\documentclass{PoS}

\title{Review of quarkonium production: \\ status and prospects}

\ShortTitle{Review of quarkonium production}

\author{\speaker{Hee Sok Chung}
\\
        Physik-Department, Technische Universit\"at M\"unchen,
James-Franck-Str. 1, 85748 Garching, Germany\\
        E-mail: \email{heesok.chung@tum.de}}

\abstract{Production cross sections of heavy quarkonia are considered as useful
tools to study various aspects of QCD. We have entered a new era of 
quarkonium production phenomenology, with the help of new measurements from the 
LHC giving access to more varieties of observables, and recent theoretical 
developments that provided us a better understanding of the short-distance 
process in which quarkonia are produced. With a more concrete understanding of
perturbative corrections and a prolific collection of data, a satisfactory
description of the quarkonium production mechanism might now be within reach.
Unfortunately, the exact mechanism of 
quarkonium production still remains elusive. Even analyses based
on the same formalism can lead to different descriptions of the production
process and give contradicting predictions of processes involving heavy
quarkonia. This implies that there are much more physics yet to be understood 
and much work yet to be done in quarkonium production phenomenology.
In this paper, we review the current status of theoretical approaches
and discuss possible strategies that may improve our understanding of heavy
quarkonium production.}

\FullConference{XIII Quark Confinement and the Hadron Spectrum - Confinement2018\\
		31 July - 6 August 2018\\
		Maynooth University, Ireland}

\begin{document}

\section{Introduction}

The many faces of QCD not only provide us a rich collection of phenomena to be 
understood but also a multitude of useful tools to study them. 
Hard processes producing heavy quarkonia, which are bound states of a heavy
quark $Q$ and a heavy antiquark $\bar Q$, belong to both categories; 
while the hard
scale lets us probe the perturbative side of QCD, the formation of heavy
quarkonium states involves physics at scales comparable or less than 
$\Lambda_{\rm QCD}$. 
Hence, heavy quarkonia are useful laboratories to study the interplay between
perturbative and nonperturbative aspects of QCD~\cite{Brambilla:2004wf, 
Brambilla:2010cs}. Moreover, many heavy quarkonia 
have clean decay channels that enable precision measurements in colliders. 
Especially, the heavy quarkonium states with $J^{PC} = 1^{--}$ such as the
$J/\psi$, $\psi(2S)$, and the $\Upsilon(nS)$ can decay into lepton pairs. 
A good understanding of the heavy quarkonium production mechanism is also 
important in other areas of QCD, such as the study of the quark-gluon plasma, 
where the $J/\psi$ 
production cross sections in heavy ion collisions are used to probe the hot and
dense phase of QCD~\cite{Matsui:1986dk}.
The production mechanism of heavy quarkonia even has important 
phenomenological implications for beyond the standard model physics models 
with strongly interacting bound states~\cite{Drees:1993uw}. 

A heavy quarkonium production process involves several distinguishable scales:
there are the perturbative scales, which are the scale of the hard process and 
the scale of the heavy quark mass $m$, and there are the scales where QCD turns
nonperturbative, which are the momentum $mv$ and the kinetic energy $mv^2$ 
of the heavy quark inside the heavy quarkonium. Here, $v$ is the typical
velocity of the heavy quark inside the meson in the meson rest frame. 
For charmonium, $v^2 \approx 0.3$, while for bottomonium, $v^2 \approx 0.1$.
When physical processes involve more than one scale that are strongly 
separated, the appropriate tool to study such processes are effective field 
theories. 
By integrating out the scales above $mv$, we obtain nonrelativistic QCD 
(NRQCD)~\cite{Caswell:1985ui, Bodwin:1994jh}. 
In NRQCD, the nonperturbative physics of scales $mv$ and $mv^2$ are
contained in the NRQCD long-distance matrix elements (LDMEs), while the
perturbative physics involving scales $m$ and above are captured in the
corresponding short-distance coefficients (SDCs). 

In NRQCD, the inclusive production cross section of a heavy quarkonium $H$ is 
given by the factorization formula~\cite{Bodwin:1994jh}
\begin{equation}
\label{eq:NRQCD-fac}%
\sigma_H = \sum_n \sigma_{Q \bar Q(n)} 
\langle 0 | {\cal O}^H(n) | 0 \rangle, 
\end{equation}
where $\sigma_{Q \bar Q(n)}$ is the SDC given by the
inclusive production cross section of a perturbative $Q \bar Q$ state in a
specific color and angular momentum state $n$, 
and $\langle 0 | {\cal O}^H(n) | 0 \rangle$ is the NRQCD LDME that governs the
evolution of the $Q \bar Q$ in the color and angular momentum state $n$ 
into the heavy quarkonium $H$. The LDMEs have known scalings with $v$, and the
sum over $n$ can be organized in powers of $v$. In practice, the sum is
truncated at a desired accuracy in $v$. 
If NRQCD factorization holds in the form of Eq.~(\ref{eq:NRQCD-fac}), the LDMEs
are universal, process-independent quantities that depend only on the
properties of the quarkonium $H$. 
A theoretical prediction of a quarkonium production cross section requires 
perturbative calculations of the SDCs and nonperturbative determinations of the
LDMEs. There have been great effort to make theoretical predictions for the
production of $J/\psi$, partly due to the availability of experimental data
over a wide range of kinematical configurations (see, for example,
Ref.~\cite{Brambilla:2010cs} and references therein). 

In present-day phenomenology of $J/\psi$ production, the factorization formula 
is usually truncated at relative order $v^4$. At leading order in $v$, the
factorization formula involves the production of $c \bar c$ in the
color-singlet, spin-triplet $S$-wave state (${}^3S_1^{[1]}$). Through order
$v^4$, the factorization formula also involves the $c \bar c$ in the 
color-octet spin-singlet $S$-wave state (${}^1S_0^{[8]}$), 
in the color-octet spin-triplet $S$-wave state (${}^3S_1^{[8]}$), 
and in the color-octet spin-triplet $P$-wave states (${}^3P_J^{[8]}$) with
$J=0,1,2$. 
Due to the approximate heavy-quark spin symmetry of NRQCD, the three
color-octet $P$-wave LDMEs 
$\langle 0 | {\cal O}^{J/\psi} ({}^3P_J^{[8]}) | 0 \rangle$ 
can be given in terms of 
$\langle 0 | {\cal O}^{J/\psi} ({}^3P_0^{[8]}) | 0 \rangle$. 
Therefore, to relative order-$v^4$ accuracy, the production cross section of
$J/\psi$ involves four NRQCD LDMEs and the corresponding SDCs. 
While the color-singlet LDME 
$\langle 0 | {\cal O}^{J/\psi} ({}^3S_1^{[1]}) | 0 \rangle$ can be obtained
from decay rates of $J/\psi$, computed from potential models, or measured using
lattice QCD, it is not yet known how to compute the color-octet LDMEs from 
first principles. Hence, the color-octet LDMEs are usually determined by
comparing Eq.~(\ref{eq:NRQCD-fac}) with data. 

Inclusive production cross section of $J/\psi$ have been measured in many
experiments. For phenomenological studies, single $J/\psi$ inclusive 
cross sections as functions of the transverse momentum $p_T$ are preferred 
because a positive definite $p_T$ gives rise to a natural choice of the hard 
scale as $p_T$ or $m_T = \sqrt{p_T^2 + m_{J/\psi}^2}$, where $m_{J/\psi}$ is 
the $J/\psi$ mass. 
The measured $p_T$-differential cross sections include the processes 
$e^+ e^- \to J/\psi +X$ by the Belle
experiment at KEKB~\cite{Pakhlov:2009nj}, 
$e^+ e^- \to e^+ e^- + J/\psi +X$ (two-photon scattering) 
by DELPHI at LEP II~\cite{Abdallah:2003du}, 
$e p \to J/\psi +X$ (photoproduction) by H1 and Zeus at 
HERA~\cite{Chekanov:2002at, Adloff:2002ex, Aaron:2010gz}, 
and the hadroproduction 
$p \bar p \to J/\psi +X$ by CDF at Tevatron~\cite{Abe:1997jz, Acosta:2004yw}, 
$p p \to J/\psi+X$ by PHENIX and STAR at RHIC~\cite{Adare:2009js,
Abelev:2009qaa},
and by CMS~\cite{Khachatryan:2010yr, Chatrchyan:2011kc, Khachatryan:2015rra,
Sirunyan:2017qdw}, 
ATLAS~\cite{Aad:2011sp, Aad:2015duc}, ALICE~\cite{Scomparin:2011zzb, 
Aamodt:2011gj, Abelev:2012kr, Abelev:2012gx},
and LHCb~\cite{Aaij:2011jh, Aaij:2012asz, Aaij:2013yaa} experiments at the LHC. 
The SDCs for these processes have been computed to next-to-leading order (NLO)
in $\alpha_s$~\cite{Ma:2010jj, Butenschoen:2010rq, Ma:2010yw, Chao:2012iv,
Butenschoen:2012px, Gong:2012ug, Butenschoen:2011yh, Butenschoen:2009zy,
Butenschoen:2011ks}. 
Currently, there are a number of determinations of $J/\psi$ LDMEs 
available in literature that are based on the NLO SDCs. 
Among them, we introduce and compare four representative examples 
from Refs.~\cite{Gong:2012ug, Butenschoen:2011yh, Shao:2014yta, Bodwin:2015iua}
that are based on the cross section measurements 
in Sec.~\ref{sec:LDMEs}. 
In Sec.~\ref{sec:Comparison}, we compare the measurements and predictions of 
other observables that can serve as tests of the LDME determinations. 
We summarize in Sec.~\ref{sec:Summary}.

\section{Determination of NRQCD LDMEs from cross section data}
\label{sec:LDMEs}%

In this section, we present four representative examples of NRQCD LDME 
determinations from Refs.~\cite{Gong:2012ug, Butenschoen:2011yh, Shao:2014yta, 
Bodwin:2015iua}, which are based on the SDCs at NLO accuracy. 
In these determinations, the color-octet LDMEs have been obtained by comparing 
the NRQCD factorization formula (\ref{eq:NRQCD-fac}) with cross section 
measurements, while the color-singlet LDME have been taken from other 
determinations. Also, in these determinations, the feed-down contributions to
the $J/\psi$ cross section from decays of $\psi(2S)$, $\chi_{c1}$ and 
$\chi_{c2}$ have been taken into account. 
In Refs.~\cite{Butenschoen:2011yh, Bodwin:2015iua}, the color-singlet LDME was
taken to be the value 
$\langle 0 |{\cal O}^{J/\psi} ({}^3S_1^{[1]})|0\rangle = 1.32$~GeV${}^3$
which was determined in Ref.~\cite{Bodwin:2007fz},
and in Refs.~\cite{Gong:2012ug, Shao:2014yta}, the value 
$\langle 0 |{\cal O}^{J/\psi} ({}^3S_1^{[1]})|0\rangle
= 1.16$~GeV${}^3$ from Ref.~\cite{Eichten:1995ch} was used. 
The difference between the two values is insignificant compared to the 
uncertainties in the SDC for the color-singlet channel, and hence, it is fair 
to say that the employed values of the color-singlet LDME are consistent with 
each
other. Meanwhile, there are considerable differences in the color-octet LDMEs. 
We show the color-octet LDMEs from Refs.~\cite{Gong:2012ug, Butenschoen:2011yh, 
Bodwin:2015iua} in Table~\ref{table1}, and present the color-octet LDMEs from 
Ref.~\cite{Shao:2014yta} below. 

In Ref.~\cite{Shao:2014yta}, only certain linear combinations of the
color-octet LDMEs could be determined, and the authors of 
Ref.~\cite{Shao:2014yta} determined ranges of the LDMEs 
by assuming that all 3 color-octet LDMEs are positive. This leads to 
$0 < \langle 0 |{\cal O}^{J/\psi} ({}^1S_0^{[8]})|0\rangle < (7.4 \pm 1.9)
\times 10^{-2}$~GeV${}^3$, 
$(0.05 \pm 0.02) \times 10^{-2}$~GeV${}^3$ 
$<\langle 0 |{\cal O}^{J/\psi} ({}^3S_1^{[8]})|0\rangle < 
(1.11 \pm 0.27) \times 10^{-2}$~GeV${}^3$, and 
$0 < \langle 0 |{\cal O}^{J/\psi} ({}^3P_0^{[8]})|0\rangle < 
(4.27 \pm 1.10) \times 10^{-2}$~GeV${}^5$. 
The values of these LDMEs are correlated in a way that when 
$\langle 0 |{\cal O}^{J/\psi} ({}^3P_0^{[8]})|0\rangle$ reaches its maximum
value, 
$\langle 0 |{\cal O}^{J/\psi} ({}^3S_1^{[8]})|0\rangle$ is also maximized, 
while the LDME $\langle 0 |{\cal O}^{J/\psi} ({}^1S_0^{[8]})|0\rangle$ becomes 
zero. In the opposite limit where 
$\langle 0 |{\cal O}^{J/\psi} ({}^3P_0^{[8]})|0\rangle$ vanishes,
$\langle 0 |{\cal O}^{J/\psi} ({}^1S_0^{[8]})|0\rangle$ reaches its maximum,
while $\langle 0 |{\cal O}^{J/\psi} ({}^3S_1^{[8]})|0\rangle$ is minimized. 
Note that the assumption in Ref.~\cite{Shao:2014yta} that all LDMEs are
positive is in tension with the determinations in 
Refs.~\cite{Gong:2012ug, Butenschoen:2011yh, Bodwin:2015iua}, where one or more
of the color-octet LDMEs are negative.

\begin{table}
  \centering
\begin{tabular}[t]{c|ccc}
& 
$\langle 0 |{\cal O}^{J/\psi} ({}^1S_0^{[8]})|0\rangle$~(GeV${}^3$) & 
$\langle 0 |{\cal O}^{J/\psi} ({}^3S_1^{[8]})|0\rangle$~(GeV${}^3$) & 
$\langle 0 |{\cal O}^{J/\psi} ({}^3P_0^{[8]})|0\rangle$~(GeV${}^5$)
\\ \hline
Ref.~\cite{Gong:2012ug} & 
$(9.7 \pm 0.9) \times 10^{-2}$ & 
$(-0.46 \pm 0.13) \times 10^{-2}$ &
$(-2.1 \pm 0.6) \times 10^{-2}$
\\ \hline
Ref.~\cite{Butenschoen:2011yh} & 
$(3.04 \pm 0.35) \times 10^{-2}$ & 
$(1.68 \pm 0.46) \times 10^{-3}$ & 
$(-9.08 \pm 1.61) \times 10^{-3}$ 
\\ \hline
Ref.~\cite{Bodwin:2015iua} & 
$(1.10\pm 0.14) \times 10^{-1}$ &
$(-7.13 \pm 3.64) \times 10^{-3}$ &
$(-7.03 \pm 3.40) \times 10^{-3}$ 
\end{tabular}
\caption{
\label{table1}%
Various determinations of color-octet NRQCD LDMEs for $J/\psi$.}
\end{table}

The determination in Ref.~\cite{Butenschoen:2011yh} is a global fit of a number
of measurements, including hadroproduction from 
PHENIX~\cite{Adare:2009js} at RHIC, 
CDF at Tevatron I~\cite{Abe:1997jz} and Tevatron II~\cite{Acosta:2004yw}, 
CMS~\cite{Khachatryan:2010yr}, ATLAS~\cite{Aad:2011sp}, 
ALICE~\cite{Scomparin:2011zzb}, and LHCb~\cite{Aaij:2011jh} at the LHC, 
photoproduction from ZEUS~\cite{Chekanov:2002at} and H1~\cite{Adloff:2002ex}
at HERA I and H1~\cite{Aaron:2010gz} at HERA II, the two-photon scattering from 
DELPHI~\cite{Abdallah:2003du} at LEP II, and 
$e^+ e^-$ annihilation from Belle~\cite{Pakhlov:2009nj} at KEKB.
The authors of Ref.~\cite{Butenschoen:2011yh} excluded data with $p_T < 1$~GeV
for photoproduction and two-photon scattering, and data with $p_T < 3$~GeV for
hadroproduction since the authors determined that the perturbative calculations 
of the SDCs are not reliable in those kinematical ranges. 
With the exception of hadroproduction, for most of the measurements considered
in Ref.~\cite{Butenschoen:2011yh}, 
the $p_T$ of $J/\psi$ does not exceed $10$~GeV. 

On the other hand, the LDME determinations in 
Refs.~\cite{Gong:2012ug, Shao:2014yta, Bodwin:2015iua} are based on the
hadroproduction data only. 
The LDME determination in Ref.~\cite{Gong:2012ug} is based on the
CDF~\cite{Acosta:2004yw} and LHCb~\cite{Aaij:2011jh} cross section
measurements.
In Ref.~\cite{Shao:2014yta}, the $J/\psi$ LDMEs were estimated from the CDF
data~\cite{Acosta:2004yw}.
The extraction in Ref.~\cite{Bodwin:2015iua} used CDF~\cite{Acosta:2004yw}
and CMS~\cite{Chatrchyan:2011kc, Khachatryan:2015rra} data.
The authors of Ref.~\cite{Bodwin:2015iua} included, in addition to the SDCs at
NLO accuracy, the leading-power (LP) fragmentation corrections 
including the leading logarithms of $p_T/m_c$ resummed to all orders in
$\alpha_s$ that have significant impact on the shape of the SDCs in $p_T$. 
Here, $m_c$ is the mass of the charm quark. 
In Refs.~\cite{Gong:2012ug, Shao:2014yta}, the data with $p_T < 7$~GeV 
was not considered in fit, whereas in Ref.~\cite{Bodwin:2015iua}, only
data with $p_T > 10$~GeV was used. These $p_T$ cuts are considerably higher
than what was employed in Ref.~\cite{Butenschoen:2011yh}, and 
removes most of the photoproduction data and data from lepton colliders from 
consideration. 
It has been found that the LDMEs extracted from high-$p_T$ hadroproduction data
are in conflict with the H1 photoproduction data and the Belle 
data~\cite{Butenschoen:2012qr, Bodwin:2015yma}.

The hadroproduction data employed in Refs.~\cite{Gong:2012ug, Shao:2014yta, 
Bodwin:2015iua} are over wide kinematical ranges, where
the $p_T$ of $J/\psi$ can even exceed 100~GeV. Since the $p_T$
cuts employed in Refs.~\cite{Gong:2012ug, Shao:2014yta, Bodwin:2015iua} are 
larger than $m_{J/\psi}$, we can understand the shape of the
$p_T$-differential cross section from its expansion in powers of $1/p_T$. 
The LP contribution,
which scales like $d \sigma^{\rm LP}/dp_T^2 \sim 1/p_T^4$, is given by
the LP fragmentation, where a single energetic parton produced in a hard
process evolves into a quarkonium~\cite{Collins:1981uw}. 
The next-to-leading power (NLP) contribution,
which scales like $d \sigma^{\rm NLP}/dp_T^2 \sim 1/p_T^6$, is given by
NLP fragmentation, where a pair of energetic partons produced in a hard process
evolves into a quarkonium~\cite{Kang:2011mg, Fleming:2012wy, Kang:2014tta}. 
The LP (NLP) fragmentation contribution is given by the convolution of the
hard part and the single(double)-parton fragmentation function. 
The SDCs can also be approximated by linear combinations
of the LP and the NLP fragmentation contributions, where the fragmentation
functions govern the evolution of the partons into the $Q \bar Q$ of specific
color and angular momentum states~\cite{Ma:2014svb}. 
In the ${}^1S_0^{[8]}$ and ${}^3P_0^{[8]}$ channels, 
the SDCs at leading order in $\alpha_s$ do not contain 
LP fragmentation contributions, and in those channels, the LP 
fragmentation contributions first appear at NLO in
$\alpha_s$~\cite{Bodwin:2014gia}. 
Hence, the NLO correction to the SDCs are enhanced by $p_T^2/m_c^2$, 
and the NLO $K$ factors are large and depend strongly on $p_T$. 
However, since the NLO SDCs already contain LP fragmentation
contributions, corrections of higher orders in $\alpha_s$ can no longer receive 
enhancements from powers of $p_T/m_c$. 

It is well known that at large $p_T$, the contribution from the ${}^3S_1^{[1]}$
channel severely underestimates the hadroproduction data~\cite{Braaten:1994vv}. 
This can also be understood from the
expansion in powers of $1/p_T$. The ${}^3S_1^{[1]}$ channel does not have a LP
contribution until NLO in $\alpha_s$. However, even at NLO, the LP
fragmentation contribution only involves the fragmentation of a charm quark. In
the hard process, the production rate of an energetic charm quark is suppressed
compared to the production rate of a gluon or a light quark. Hence, it is not
until NNLO in $\alpha_s$ that the ${}^3S_1^{[1]}$ channel receives a sizable
contribution from LP fragmentation. 
On the other hand the color-octet channels receive LP fragmentation
contributions from LO (${}^3S_1^{[8]}$ channel) and NLO (${}^1S_0^{[8]}$ and
${}^3P_0^{[8]}$ channels), and when compared to the color-singlet channel,
the suppression from powers of $v$ can be overcome by enhancement from inverse
powers of $\alpha_s$ at the scale of $p_T$. In all LDME determinations in 
Refs.~\cite{Gong:2012ug, Butenschoen:2011yh, Shao:2014yta, Bodwin:2015iua},
the contributions from the color-octet channels dominate the hadroproduction
rate. 

The color-octet LDMEs in Refs.~\cite{Gong:2012ug,Bodwin:2015iua} have
uncertainties that are strongly correlated. In Ref.~\cite{Shao:2014yta}, only
two linear combinations of the three color-octet LDMEs could be determined, and
the ranges of the LDMEs were constrained by requiring the positivity of all
three color-octet LDMEs. These imply that hadroproduction data cannot 
constrain all three color-octet LDMEs strongly. The origin of this problem is 
the fact that the three SDCs corresponding to the three color-octet LDMEs 
are approximately linearly dependent~\cite{Ma:2010yw}. 
Since each SDC can also be approximated by a linear combination
of the LP and the NLP fragmentation contributions,
there is always an approximate linear dependence between three SDCs at large
$p_T$. In the next section, we will discuss the efforts to further constrain 
the ranges of the matrix elements from Ref.~\cite{Shao:2014yta} by using other 
measurements as constraints. 

It is worth noting that, in the LDME extractions in Refs.~\cite{Gong:2012ug,
Shao:2014yta, Bodwin:2015iua}, the color-octet LDMEs 
$\langle 0 |{\cal O}^{J/\psi} ({}^3S_1^{[8]})|0\rangle$ and 
$\langle 0 |{\cal O}^{J/\psi} ({}^3P_0^{[8]})|0\rangle$ have same signs. 
The SDCs for the ${}^3S_1^{[8]}$ and ${}^3P_0^{[8]}$ channels have 
large leading-power
contributions compared to the next-to-leading power contributions, and the
shapes of the SDCs in $p_T$ are 
incompatible with hadroproduction data. Therefore it is necessary that the
either the sum of contributions from the two channels are small, or there
should be large cancellations between the contributions in order to describe 
the large-$p_T$ data. Since the SDCs for the ${}^3S_1^{[8]}$ and 
${}^3P_0^{[8]}$ channels have opposite signs at large $p_T$, for cancellations
to occur, the corresponding LDMEs need to have same signs as in
Refs.~\cite{Gong:2012ug, Shao:2014yta, Bodwin:2015iua}.
On the other hand, the color-octet LDMEs $\langle 0 |{\cal O}^{J/\psi}
({}^3S_1^{[8]})|0\rangle$ and
$\langle 0 |{\cal O}^{J/\psi} ({}^3P_0^{[8]})|0\rangle$ have opposite signs in
Ref.~\cite{Butenschoen:2011yh}, and so, the contributions from those channels
add at large $p_T$. 
These qualitative differences in the color-octet LDMEs have interesting 
implications in the prediction of other observables. 
For example, for hadroproduction and photoproduction 
at $p_T$ comparable to $m_{J/\psi}$, the SDCs 
for the ${}^3S_1^{[8]}$ and ${}^3P_0^{[8]}$ channels are both positive. 
Hence, positive color-octet LDMEs $\langle 0 |{\cal O}^{J/\psi}
({}^3S_1^{[8]})|0\rangle$ and 
$\langle 0 |{\cal O}^{J/\psi} ({}^3P_0^{[8]})|0\rangle$ can enhance the cross
section when $p_T$ is comparable to $m_{J/\psi}$, and lead to overestimation of
the cross section~\cite{Butenschoen:2012qr}. 
Comparison between the predictions of LDME determinations for other observables 
such as the polarization of $J/\psi$ produced in hadron colliders will be
discussed in the next section.

\section{Comparison with other observables}
\label{sec:Comparison}%

The large number of $J/\psi$ produced in hadron colliders give us access to
observables other than the production rate. These observables can be sensitive
to the color-octet LDMEs in a different combination than the cross sections,
and can provide independent tests of the color-octet LDMEs. 
In this section we consider three such examples that have been measured at the
LHC : the polarization of $J/\psi$, the momentum distribution of $J/\psi$ 
inside a jet, and the production rate of $\eta_c$.

\subsection{Polarization of $J/\psi$}

The polarization of $J/\psi$ can be measured from the angular distribution of
the muon pairs from the decay of $J/\psi$ in the meson rest frame. 
The axis that defines the polarization of $J/\psi$ is usually chosen to be the
direction of the boost from the lab frame to the rest frame of the $J/\psi$ : 
such choice is called the center-of-mass helicity frame. 
The polarization of $J/\psi$ have been measured by CDF~\cite{Affolder:2000nn,
Abulencia:2007us} at Tevatron, and by 
ALICE~\cite{Abelev:2011md, Acharya:2018uww},
CMS~\cite{Chatrchyan:2013cla}, and
LHCb~\cite{Aaij:2013nlm, Aaij:2017egv} at the LHC. 
While the CDF Run I~\cite{Affolder:2000nn} and Run II~\cite{Abulencia:2007us} 
measurements are incompatible with each other, 
the measurements at the LHC are in reasonable agreement and imply that the 
$J/\psi$'s produced in $pp$ collisions are almost unpolarized. 

The polarization of $J/\psi$ is sensitive to the color-octet LDMEs. 
At large $p_T$, the $J/\psi$ produced in the 
${}^3S_1^{[8]}$ and ${}^3P_0^{[8]}$ channels have strong transverse
polarization, while the ${}^1S_0^{[8]}$ channel produces 
unpolarized $J/\psi$, because the ${}^1S_0^{[8]}$ state is isotropic. 
Hence, in order to produce unpolarized $J/\psi$, either the ${}^1S_0^{[8]}$ 
channel should dominate the cross section, or the contributions form 
${}^3S_1^{[8]}$ and ${}^3P_0^{[8]}$ channels must have large cancellations so
that the production rate of transversely polarized $J/\psi$ is reduced. 
Hence, the color-octet LDMEs determined in Refs.~\cite{Gong:2012ug,
Shao:2014yta, Bodwin:2015iua}, which feature large cancellations between the
contributions from the ${}^3S_1^{[8]}$ and ${}^3P_0^{[8]}$ channels in the
cross section, lead to predictions of small or almost vanishing $J/\psi$ 
polarization at large $p_T$. In contrast, the
color-octet LDMEs determined in Ref.~\cite{Butenschoen:2011yh} lead to
a prediction of transversely polarized $J/\psi$ at large $p_T$, because in
Ref.~\cite{Butenschoen:2011yh}, the
contributions from ${}^3S_1^{[8]}$ and ${}^3P_0^{[8]}$ channels 
add and enhance the transverse cross 
section at large $p_T$~\cite{Butenschoen:2012px, Butenschoen:2012qr, 
Butenschoen:2012qh}. 
This prediction does not agree with measurements at the 
LHC~\cite{Chatrchyan:2013cla}. 

In Ref.~\cite{Chao:2012iv}, the authors presented a set of color-octet LDMEs 
that were constrained by fitting the high-$p_T$ CDF cross section and 
polarization data simultaneously. This simultaneous fit leads to a large 
$\langle 0 |{\cal O}^{J/\psi} ({}^1S_0^{[8]})|0\rangle$, while 
$\langle 0 |{\cal O}^{J/\psi} ({}^3S_1^{[8]})|0\rangle$ and 
$\langle 0 |{\cal O}^{J/\psi} ({}^3P_0^{[8]})|0\rangle$ are small and positive. 
This result is compatible with the ranges of color-octet LDMEs obtained by the
same authors in Ref.~\cite{Shao:2014yta}, while 
having much narrower ranges, and therefore the LDMEs in 
Ref.~\cite{Chao:2012iv} can lead to more concrete predictions of 
observables. These LDMEs predict slightly longitudinal $J/\psi$ at the LHC,
which is in tension with Ref.~\cite{Shao:2014yta}, which predicts slightly
transverse $J/\psi$ at the LHC. 

\subsection{Momentum distribution of $J/\psi$ in jet}

Another useful observable is the momentum distribution of a $J/\psi$ in a
jet~\cite{Baumgart:2014upa}. 
The momentum distribution can be measured as a function of $z$, which is
the fraction of the momentum of $J/\psi$ compared to the momentum of the jet. 
The shape of this distribution is sensitive to the color-octet LDMEs;
the ${}^3S_1^{[8]}$ and ${}^3P_0^{[8]}$ channels have distributions that rise
as $z \to 1$, while in the ${}^1S_0^{[8]}$ channel, the distribution falls as
$z \to 1$. 
So far, the LHCb~\cite{Aaij:2017fak} and CMS~\cite{CMS:2018mjn} 
have measured the momentum distribution 
of $J/\psi$ in jet. Both measurements have a common feature that the
distribution falls as $z \to 1$. 
Since ${}^3S_1^{[8]}$ and ${}^3P_0^{[8]}$ channels have rising distributions
as $z \to 1$, in order to have a falling distribution as $z \to 1$ 
we need a large cancellation between the ${}^3S_1^{[8]}$ and ${}^3P_0^{[8]}$
channels, or the ${}^1S_0^{[8]}$ channel must dominate the cross section. 
In Ref.~\cite{Bain:2017wvk}, the authors presented a calculation of the
momentum distribution of $J/\psi$ in jet based on the fragmenting jet functions 
(FJF) formalism developed in Ref.~\cite{Procura:2009vm}. 
The results from the FJF formalism are found to agree with calculations based 
on PYTHIA with modifications to properly accommodate the gluon fragmentation 
process~\cite{Bain:2017wvk, Bain:2016clc}. 
The authors of Ref.~\cite{Bain:2017wvk} made predictions based on
three independent determinations of color-octet LDMEs based on the NLO
calculation of the SDCs : 
LDME set 1 is from Refs.~\cite{Butenschoen:2011yh, Butenschoen:2012qr},
which is determined from the global fit of $J/\psi$ inclusive cross sections 
without considering feed-down contributions. 
LDME set 2 is from Ref.~\cite{Chao:2012iv}, where the LDMEs were obtained by 
fitting the high-$p_T$ CDF cross section and polarization data simultaneously. 
LDME set 2 is compatible with Ref.~\cite{Shao:2014yta}. 
Finally, LDME set 3 from Ref.~\cite{Bodwin:2014gia}, where the LDMEs were 
obtained from high-$p_T$ cross section data from CDF and CMS, 
were also considered. 
The LDMEs in set 3 are compatible with those in Ref.~\cite{Bodwin:2015iua}. 
Recall that, in Refs.~\cite{Ma:2010yw, Shao:2014yta, Bodwin:2015iua}, there
are large cancellations between the ${}^3S_1^{[8]}$ and ${}^3P_0^{[8]}$
channels, and in Ref.~\cite{Bodwin:2015iua}, the ${}^1S_0^{[8]}$ channel
dominates the hadroproduction cross section. 
The same holds true for LDME sets 2 and 3, so that the
resulting distributions fall as $z \to 1$; the overall shapes of the
distributions are in reasonable agreement with data~\cite{Bain:2017wvk}. 
On the other hand, it was shown in Ref.~\cite{Bain:2017wvk} that the LDME set 
1 gives a much flatter distribution that disagrees with data. 
It has not been reported whether the full ranges of color-octet LDMEs from
Ref.~\cite{Shao:2014yta} lead to a concrete prediction of the $J/\psi$ momentum
distribution in jet.

\subsection{Production rate of $\eta_c$}

Due to the approximate heavy-quark spin symmetry of NRQCD, the LDMEs for the
production of $J/\psi$ can be related with the LDMEs for the production of
$\eta_c$~\cite{Bodwin:1994jh}. 
Hence, a determination of $J/\psi$ LDMEs leads to prediction of the
$\eta_c$ production rate. The $p_T$-differential cross section of $\eta_c$ has 
been measured by LHCb~\cite{Aaij:2014bga}, 
and so, this measurement can serve as a test of the
$J/\psi$ production mechanism. 

The $\eta_c$ production cross section involves at leading order in $v$ the
color-singlet spin-singlet channel (${}^1S_0^{[1]}$), and through relative
order $v^4$ the three color-octet channels ${}^3S_1^{[8]}$, 
${}^1P_1^{[8]}$, and ${}^1S_0^{[8]}$. 
The corresponding LDMEs can be obtained from $J/\psi$ LDMEs that differ by one
unit of spin, with appropriate conversion factors obtained from spin 
multiplicities of the LDMEs. On the other hand, the behavior of the SDCs are
different from the case of $J/\psi$ hadroproduction; 
for $\eta_c$ hadroproduction
at large $p_T$, the SDC for the color-singlet channel does not have the 
same strong suppression from $\alpha_s$ as the color-singlet SDC for the
$J/\psi$, 
and even the contribution from the color-singlet channel is large 
enough to fully accommodate the $\eta_c$ hadroproduction 
data~\cite{Butenschoen:2014dra}. 
Moreover, the SDCs for the color-octet channels are 
positive for all channels, and so, the color-octet LDMEs that lead to 
large cancellations in the $J/\psi$ production rate can give large enhancements
to the $\eta_c$ production rate. Also, since the SDC for the ${}^3S_1^{[8]}$
channel is much larger than the SDC for the ${}^1S_0^{[8]}$ channel, if the
LDME $\langle 0 |{\cal O}^{J/\psi} ({}^1S_0^{[8]})|0\rangle \approx
\langle 0 |{\cal O}^{\eta_c} ({}^3S_1^{[8]})|0\rangle$ is large, 
the contribution from the ${}^3S_1^{[8]}$ channel to the $\eta_c$ cross section
can be much larger than the contribution from the ${}^1S_0^{[8]}$ channel to
the $J/\psi$ cross section. Hence, if the $J/\psi$ production cross section
is dominated by the ${}^1S_0^{[8]}$ channel, the $\eta_c$ cross section will be
strongly enhanced by the ${}^3S_1^{[8]}$ channel. 

The authors of Ref.~\cite{Butenschoen:2014dra} found that, the LDMEs from the
global fit of $J/\psi$ cross sections~\cite{Butenschoen:2011yh}, 
the LDMEs from simultaneous fit to production rate and polarization of $J/\psi$ 
at the Tevatron~\cite{Chao:2012iv}, and the LDMEs from $J/\psi$ hadroproduction 
data~\cite{Gong:2012ug, Bodwin:2014gia} lead to predictions that overestimate 
the $\eta_c$ cross section data. The LDMEs from Ref.~\cite{Shao:2014yta},
which were obtained from the CDF $J/\psi$ cross section data with the positivity
assumption, is found to be compatible with data~\cite{Han:2014jya,
Zhang:2014ybe}. In Refs.~\cite{Han:2014jya, Zhang:2014ybe}, 
it has been shown to be possible to reduce the ranges of the color-octet LDMEs
in Ref.~\cite{Shao:2014yta} by using the $\eta_c$ cross section data. 
However, the LDMEs determined in Refs.~\cite{Han:2014jya, Zhang:2014ybe} 
have much smaller values of 
$\langle 0 |{\cal O}^{J/\psi} ({}^1S_0^{[8]})|0\rangle$, and 
are in conflict with the LDMEs from Ref.~\cite{Chao:2012iv} determined from
the CDF cross section and polarization data. 
Also, the authors of Ref.~\cite{Butenschoen:2014dra} have pointed out that the
prediction for the $J/\psi$ polarization based on the LDMEs in
Ref.~\cite{Han:2014jya} is in tension with the LHC data.

\section{Summary and outlook}

\label{sec:Summary}%

In this paper we have reviewed the current status of $J/\psi$ production
phenomenology based on the NRQCD factorization formalism. 
A prediction based on the NRQCD factorization formalism requires knowledge of
the short-distance coefficients, which are perturbatively calculable, and the
long-distance matrix elements, which are nonperturbative quantities. 
Since it is not known how to compute color-octet long-distance matrix elements,
they are usually determined phenomenologically by comparing the NRQCD 
factorization formula with measurements. We presented four 
representative examples of the long-distance matrix element determinations from
cross section measurements, based on the state-of-the-art calculations of the
short-distance coefficients at next-to-leading order accuracy. 
The long-distance matrix elements determined from different choices of data 
can disagree with one another, 
and none of the determinations are able to give a comprehensive description of 
the important observables, such as the cross section measurements at various
kinematical configurations, polarization, 
momentum distribution of $J/\psi$ in jet, and the $\eta_c$ cross section data
at the LHC, at a satisfactory level. 
It is therefore unclear whether a prediction of an observable based a specific 
determination of the long-distance matrix elements can be regarded trustworthy, 
except for cases where the observable is insensitive to specific
values of the color-octet long-distance matrix elements at the current level of
experimental accuracy (see, for example, Ref.~\cite{Feng:2018cai}). 

One may question the reliability of the perturbative expansion of the 
short-distance coefficients. At small $p_T$, $\alpha_s$ becomes large, 
while at $p_T$ much larger than $m_c$, 
large corrections enhanced by powers and logarithms of $p_T/m_c$ may appear. 
In both cases, 
the convergence of the perturbation series can be spoiled. 
However, the next-to-leading order corrections to the short-distance 
coefficients show
that, when $p_T$ is of the order of the mass of the $J/\psi$, the corrections 
are moderate in size~\cite{Butenschoen:2010rq, Ma:2010yw}. 
While the next-to-leading order corrections are large and have strong
dependencies in $p_T$ when $p_T$ is much larger than $m_c$,
these large corrections are well understood in
terms of the factorized expansion in powers of $1/p_T$~\cite{Ma:2014svb}. 
And, from the factorization theorems of perturbative QCD~\cite{Collins:1981uw}, 
we do not expect large
corrections enhanced by even more powers of $p_T/m_c$ to appear at higher 
orders in $\alpha_s$. Furthermore, it has been shown that the resummation of 
leading logarithms of $p_T/m_c$ does not impact the shape of the 
short-distance coefficients significantly within the kinematical ranges where
data is currently available~\cite{Bodwin:2015iua, Bodwin:2014gia}.
Hence, it is doubtful that higher-order corrections in $\alpha_s$ can affect 
the short-distance coefficients significantly at large $p_T$. It is still
possible that higher order corrections in $\alpha_s$ change the relative sizes
of contributions at leading power and next-to-leading power in $1/p_T$ and
change the shape of the short-distance coefficients, which would affect 
the determinations of long-distance matrix elements from
data. While a complete calculation of next-to-next-to-leading order corrections 
is currently out of reach, there have been some progress in the calculation of
the single-parton fragmentation functions for the ${}^1S_0^{[1]}$ and
${}^1S_0^{[8]}$ channels to next-to-leading order in 
$\alpha_s$~\cite{Artoisenet:2014lpa, Artoisenet:2018dbs, Zhang:2018mlo}, which
allow us to compute the next-to-next-to-leading order correction to the 
short-distance coefficients for those channels at leading power in $1/p_T$. 

Since in $J/\psi$ hadroproduction, the color-octet channels have larger 
contributions to the cross 
section than the color-singlet channel, one may also doubt the convergence of
the expansion in powers of the velocity $v$. 
Including contributions of higher orders in $v$ in the factorization formula
(\ref{eq:NRQCD-fac}) would result in a huge increase of nonperturbative
unknowns, and consequently, a loss of predictive power.
It should be noted, however, that none of the determinations of the $J/\psi$ 
long-distance matrix elements that we considered violate the velocity-power 
counting, as in all cases, the color-octet long-distance matrix elements are at 
least an order of magnitude smaller than the color-singlet matrix element,
which is consistent with $v^4 \approx 0.1$. 
The enhancement of the color-octet channel contributions is of 
dynamical origin, and is process dependent; in $J/\psi$ photoproduction or 
$\eta_c$ hadroproduction, we do not see such a dramatic suppression of the
color-singlet channel contribution. 
An alternate formulation devised to improve the convergence of the velocity 
expansion has been suggested in Ref.~\cite{Ma:2017xno}.

In Refs.~\cite{Kang:2013hta, Ma:2014mri, Ma:2018qvc}, the authors employed the 
color glass condensate framework~\cite{Kovchegov:2012mbw, Gelis:2010nm, 
Weigert:2005us, Iancu:2003xm} to describe $J/\psi$ hadroproduction for 
$p_T$ comparable to, or even smaller than, the $J/\psi$ mass, where the
short-distance process is sensitive to the proton structure at very small
Bjorken $x$. 
Based on the long-distance matrix elements determined in
Ref.~\cite{Chao:2012iv}, where all color-octet long-distance matrix elements
are positive, the authors found good agreement with LHC data for both
cross section and polarization at small $p_T$. 
This is in contrast with Ref.~\cite{Butenschoen:2011yh}, where the standard
collinear factorization was used to describe hadron collisions with parton
distribution functions; in Ref.~\cite{Butenschoen:2011yh}, the authors found
that the color-octet long-distance matrix element must have different
signs in order to describe low-$p_T$ cross section data. 
This may hint that for $J/\psi$ hadroproduction at small $p_T$, 
collinear factorization might be unsuitable for describing hadron collisions. 
It is however yet unknown whether there would be such large corrections to
photoproduction at HERA, or production processes in lepton colliders.

So far, studies of the color-octet long-distance matrix elements were limited
to phenomenological determinations. It is therefore highly desirable to have a
constraint, or even a calculation, of the long-distance matrix elements from 
first principles. Recent developments such as the possible lattice 
calculation of parton distribution functions~\cite{Ji:2013dva,
Radyushkin:2016hsy, Chambers:2017dov, Ma:2017pxb} give hope that a 
first-principles determination of nonperturbative NRQCD long-distance 
matrix elements might be achievable through a thorough investigation 
of nonrelativistic effective field theories and lattice QCD. 
For example, the potential NRQCD effective field theory~\cite{Brambilla:1999xf},
through the separation of scales $mv$ and $mv^2$, can reveal previously
unknown properties and symmetries of the NRQCD long-distance matrix elements,
and may give rise to formulations of the long-distance matrix elements that 
can be computed in lattice QCD.

\acknowledgments

I would like to thank Nora Brambilla for her valuable comments and careful
reading of the manuscript. 
This work was supported by the Alexander von Humboldt Foundation and 
the DFG cluster of excellence
`Origin and Structure of the Universe' (www.universe-cluster.de).

\end{document}